\documentclass[preprint,superscriptaddress,onecolumn,showpacs,showkeys]{revtex4-1}
\usepackage{textcomp}
\usepackage{eurosym}
\usepackage{amsfonts}
\usepackage{array}
\usepackage{amsthm}
\usepackage{bm}
\usepackage{palatino}
\usepackage[colorinlistoftodos]{todonotes}
\usepackage{mathpazo}
\usepackage{supertabular}
\usepackage{subfig}
\usepackage{amssymb}
\usepackage{eurosym}
\usepackage{amsmath}
\usepackage{epsfig}
\usepackage{graphics}
\usepackage{color}
\usepackage[font={footnotesize,it}]{caption}
\usepackage[colorlinks=true,
            linkcolor=red,
            urlcolor=blue,
            citecolor=green]{hyperref}
\usepackage{graphicx,color}
\usepackage{pdfpages}

\makeatletter
\AtBeginDocument{\let\LS@rot\@undefined}
\makeatother

\begin{document}


\setcounter{MaxMatrixCols}{10}

\title{Fermion Clouds Around $z=0$ Lifshitz Black Holes}
\author{G\"{u}lnihal Tokg\"{o}z}
\email{gulnihal.tokgoz@emu.edu.tr}
\author{\.{I}zzet Sakall{\i}}
\email{izzet.sakalli@emu.edu.tr}
\email{izzet.sakalli@gmail.com}

\affiliation{Physics Department, Faculty of Arts and Sciences, Eastern
Mediterranean University, Famagusta, North Cyprus, via Mersin 10, Turkey}

\begin{abstract}
The Dirac equation is solved in the $z=0$ Lifshitz black hole ($Z0$LBH)
spacetime. The set of equations representing the Dirac equation in the
Newman-Penrose (NP) formalism is decoupled into a radial set and an angular
set. The separation constant is obtained with the aid of the spin weighted
spheroidal harmonics. The radial set of equations, which is independent of
mass, is reduced to Zerilli equations (ZEs)\ with their associated
potentials. In the near horizon (NH) region, these equations solved in terms
of the Bessel functions of the first and second kinds arising from the
fermionic perturbation on the background geometry. For computing the BQNMs
instead of the ordinary quasinormal modes (QNMs), we first impose the purely
ingoing wave condition at the event horizon. And then, Dirichlet boundary
condition (DBC) and Newmann boundary condition (NBC) are applied in order to
get the resonance conditions. For solving the resonance conditions we follow
an iteration method. Finally, Maggiore's method (MM) is employed to derive
the entropy/area spectra of the $Z0$LBH which are shown to be equidistant.

\end{abstract}

\maketitle
\tableofcontents

\section{Introduction}

The interaction of the BHs (depending on the couplings of spin-rotation,
field mass-black hole (BH) mass, field charge- BH charge) with a fermionic
(Dirac) field has attracted deep theoretical interest and it has motivated
the researchers to have a better understanding of various physical fields,
especially the spin 1/2 particles, in the vicinity of the BHs.

The seminal works of Bekenstein and Hawking \cite{main1,main2,Beks1} has
shown that BH entropy ($S_{BH}$) so that BH area ($\mathcal{A}_{BH}$) should
be quantized in discrete and equidistant levels (see for a detailed
discussion \cite{Beks2,Beks3,Beks4,Beks5}). There is a proportionality
between $S_{BH}$ and $\mathcal{A}_{BH}$ \ ($S_{BH}=\frac{\mathcal{A}_{BH}}{4}
$), which is attested from the adiabatic invariance \cite{Efst} . It has
been proposed by Bekenstein that $\mathcal{A}_{BH}$ should have the
following discrete, equidistant spectrum (for the family of Schwarzschild
BHs \cite{Beks3,Beks4}):%
\begin{equation}
\mathcal{A}_{BH}=\epsilon \hbar n,\text{ \ \ \ \ \ \ }n=0,1,2,...
\label{1-1}
\end{equation}%
where $\epsilon $ is known as the unknown fudge factor \cite{izr10}. One
can conclude from the above equation that the minimum change in the horizon
area of the Schwarzschild BH ($\epsilon =8\pi $) \cite%
{maggiore,vagenas,medved} is $\Delta \mathcal{A}_{\min }=8\pi \hbar $. Among
the methods that have been improved to obtain the entropy/area spectra of
the numerous BHs inspired from Bekenstein (see \cite{myGua} and references
therein), MM \cite{maggiore}, which was based on Kunstatter's study \cite%
{myKuns} is in a full agreement with Bekenstein's original result (\ref{1}%
). In Kunstatter's study \cite{myKuns}, the adiabatic invariant quantity ($%
I_{adb}$) was expressed by:

\begin{equation}
I_{adb}=\int \frac{dM}{\Delta \omega },  \label{2-2}
\end{equation}

where $\Delta \omega =\omega _{n-1}-\omega _{n}$ is the transition frequency
of a BH having energy (mass) $M$. Furthermore, $I_{adb}$ was generalized to
cover the BHs \ which are massive, charged, and rotating (hairy) (see \cite%
{myKwon} and references therein) as follows :

\begin{equation}
I_{adb}=\int \frac{\mathcal{T_{H}}dS_{BH}}{\Delta \omega },  \label{3-3}
\end{equation}

in which the temperature of the BH is denoted by $\mathcal{T_{H}}$. The
adiabatic invariant quantity acts as a quantized quantity $(I_{adb}\simeq
n\hbar )$ for the highly excited states ($n\rightarrow \infty $) according
to the Bohr--Sommerfeld quantization rule \cite{BS}. Therefore, the
imaginary part of the frequency becomes dominant ($\omega _{I}\gg \omega
_{R} $), and this yields that $\Delta \omega \simeq \Delta \omega _{I}$.
Meanwhile, Hod \cite{myHod1,myHod2} was the first physicist to argue that
one can be use the QNMs \cite{Corless,Fernando} in order for computing the
transition frequency. Maggiore, inspired by Hod's suggestion, considered the
Schwarzschild BH as a highly damped harmonic oscillator and using a
different method, re-derived Bekenstein's original result (\ref{1}).
Recently, MM has been employed in numerous studies of BH quantization in the
literature (see for instance \cite%
{myLi,kothawala,Ortega,Sakalli1,Sakalli22,gul}).

In this paper,we mainly investigate the entropy/area spectra of a
four-dimensional Lifshitz BH (LBH) \cite{lu} possessing a particular dynamical
exponent $z=0$. The Lifshitz spacetimes have attracted attention from the
researchers working on condensed matter and quantum field theories \cite%
{kachru} for being invariant under anisotropic scale and characterizing
gravitational dual of strange metals \cite{hartnoll}. We start our analysis
with the calculations of its quasilocal mass $M_{QL}$ \cite{brown0} and
temperature by using the Wald's entropy \cite{Eune} and statistical
temperature formula \cite{wald0,wald1}, respectively. In fact, this problem
was previously studied \cite{gulzzlbh,catalan,xu}. The QNMs were calculated
and it was shown that \ the $Z0$LBH possesses a discrete and equidistant
spectrum under massive scalar field perturbations. In this study, we
reconsider the problem with fermionic perturbations and instead of the
ordinary QNMs, we derive the boxed QNMs (BQNMs) \cite%
{Hod,gul,BQNM1,BQNM2,BQNM3} that are the characteristic resonance spectra of
the confined scalar fields in the $Z0$LBH geometry. To this end, we consider
the Dirac equations with $q=0$ and $\mu ^{\ast }=0$ spin-$\frac{1}{2}$ test
particles in the $Z0$LBH spacetime \cite{lu}\ and we show how the separation
of the angular and the radial equations yields a Schr\"{o}dinger-like wave
equation (SLE) or the so-called Zerilli equation (ZE) \cite{chandra} with
its NH form.

To this end, we consider a mirror (confining cavity) surrounding the $Z0$LBH
which is located at a constant radial coordinate. Next, we impose that the
fermionic field should terminate at the mirror's location, which requires
two boundary conditions to be used: DBC and NBC \cite{Hod,Sakalli,okawa,gul}%
. With this scenario,we focus our analysis of the ZE in the NH region \cite%
{Hod}.

After getting the NH form of the ZE, we show that the radial equation has
the solution in terms of the Bessel functions of the first and second kind 
\cite{Abramowitz}. Finally, we use an iteration scheme and show how the
Dirac BQNMs and the quantum spectra of entropy and area of the $Z0$LBH\ are
obtained.

The present paper is organized as follows. In section 2, we review the
background geometry ($Z0$LBH) and we obtain the quasilocal mass ($M_{QL}$)
via the Brown-York (BY) formalism \cite{brown0,brown1}. In section 3, we
solve the Dirac equations for the uncharged massless spin 1/2 particles, in
the framework of NP formalism. In particular, we analyze the effective
potential for the ZE. We show the existence of the Dirac BQNMs of $Z0$LBH in
section 4. Finally the paper ends with a discussion of our findings in
section 5. Throughout this study we use units with $G=c=\hbar =1$.

\section{$Z0$LBH Spacetime}

In this section we introduce the Lifshitz spacetimes \cite{catalan,kachru}
and the special case $Z0$LBH\ in four dimensions \cite{lu}\ on which we
focus our analysis, briefly. Lifshitz spacetimes are described by the
following line element \cite{kachru,catalan}

\begin{equation}
ds^{2}=-\frac{r^{2z}}{l^{2z}}dt^{2}+\frac{l^{2}}{r^{2}}dr^{2}+\frac{r^{2}}{%
l^{2}}d\vec{x}^{2},  \label{1}
\end{equation}

where $l$ is the length scale in the geometry, $z$ is the dynamical exponent
and $\vec{x}$ is the D-2 spatial vector. The action corresponding to the
Einstein -Weyl gravity is given by \cite{lu}%
\begin{equation}
S=\frac{1}{2\kappa ^{2}}\int \sqrt{-g}d^{4}x(R-2\Lambda +\frac{1}{2}\alpha
\left\vert Weyl\right\vert ^{2}),  \label{2}
\end{equation}

where $\kappa ^{2}=8\pi G$ , $\left\vert Weyl\right\vert ^{2}=R^{\mu \nu
\rho \sigma }R_{\mu \nu \rho \sigma }-2R^{\mu \nu }R_{\mu \nu }+\frac{1}{3}%
R^{2}$\ and $\alpha =\frac{z^{2}+2z+3}{4z(z-4)}$ is a constant that
corresponds to $\alpha =\infty $ for $z=0$ or $z=4$.

It is shown that the static asymptotically LBH solutions for both $z=4$ and $%
z=0$ are received by the CG \cite{lu,cai} and also $z=3$ and $z=4$ LBH
solutions exist in the Ho\v{r}ava-Lifshitz gravity \cite{lu,catalan,herrera}.
Now we turn our attention to a special case of CG, the four dimensional $Z0$%
LBH, with its metric given by \cite{lu}%
\begin{equation}
ds^{2}=f\left( r\right) dt^{2}-\frac{4}{r^{2}f\left( r\right) }%
dr^{2}-r^{2}d\Omega _{2,k}^{2}  \label{3}
\end{equation}

with \cite{herrera}

\begin{equation}
d\Omega _{2,k}^{2}=%
\begin{array}{c}
d\theta ^{2}+d\varphi ^{2},\text{ \ \ \ \ \ \ \ \ \ \ \ \ \ \ \ \ }k=0 \\ 
d\theta ^{2}+\sin ^{2}\left( \theta \right) d\varphi ^{2},\text{ \ \ \ \ \ \ 
}k=1 \\ 
d\theta ^{2}+\sinh ^{2}\left( \theta \right) d\varphi ^{2},\text{ \ \ }k=-1.%
\end{array}
\label{4}
\end{equation}

The metric function $f(r)$ is defined as%
\begin{equation}
f\left( r\right) =1+\frac{c}{r^{2}}+\frac{c^{2}-k^{2}}{3r^{4}}  \label{5}
\end{equation}

In this metric $d\Omega _{2,k}^{2}$ becomes 2-D sphere in accordance with
our choice $k=1$. One may choose $k=-1$ and get the unit hyperbolic plane or 
$k=0$ to get 2-torus \cite{herrera}. The metric given in Eq. (\ref{3})
conformally describes an AdS BH if one sets $k=-1$ ,on the other hand, it
describes a dS BH if one choose $k=1$ \cite{lu}. The solution has a
singularity at $r=0$ however this singularity becomes naked for $k=0$. There
is an horizon for $k=\pm 1$ solution of the form \cite{lu,herrera}%
\begin{equation}
r_{h}^{2}=\frac{1}{6}\left( \sqrt{3(4-c^{2})}-3c\right) .  \label{6}
\end{equation}

Since $r_{h}^{2}$ is positive ($r_{h}^{2}=1$), it is required to be $-2\leq
c<1$. Our focus is on $k=1$ solution with $c=-1$ ,so now the metric is given
by%
\begin{equation}
ds^{2}=f\left( r\right) dt^{2}-\frac{4}{r^{2}f\left( r\right) }dr^{2}-r^{2}%
\left[ d\theta ^{2}+\sin ^{2}\left( \theta \right) d\varphi ^{2}\right] ,
\label{7a}
\end{equation}

with

\begin{equation}
f\left( r\right) =1-\frac{r_{h}^{2}}{r^{2}}.  \label{7b}
\end{equation}

The corresponding Ricci scalar and Kretschmann scalar change in the
following proportions at spatial infinity, respectively%
\begin{eqnarray}
R &=&R_{\mu }^{\mu }\sim \frac{5(c^{4}-2c^{2}+1)}{12r^{8}},  \nonumber \\
K &=&R^{\mu \nu \rho \sigma }R_{\mu \nu \rho \sigma }\sim \frac{%
25(c^{4}-2c^{2}+1)}{12r^{8}}.  \label{8}
\end{eqnarray}

\subsection{Mass Computation of $Z0$LBH via BY Formalism}

The spacetime ,as a union of space, time and gravitation in GR, has a
curvature singularity which stands for the gravitational force. In GR,
energy conservation is fundamental. Our spacetime is a four dimensional
manifold,so that, one should be able to get the information from 2-D
boundary surface $d\Omega _{2,k}^{2}$\ (hypersurface $\Sigma $). Quasilocal
mass $M_{QL}$\ \ \cite{brown0,brown1}\ measures the density of matter.

We will consider\ BY formalism, in order to calculate $M_{QL}$. In this
formalism, a spherically symmetric four dimensional metric solution is given
by \cite{brown0,brown1}%
\begin{equation}
ds^{2}=-F(R)^{2}dt^{2}+\frac{dR^{2}}{G(R)^{2}}+R^{2}d\Omega _{2}^{2},
\label{9}
\end{equation}

which adopts the $M_{QL}$ with the following definition \cite%
{mazhari0,mazhari1}%
\begin{equation}
M_{BY}=\frac{N-2}{2}R^{N-3}F(R)\left\{ G_{ref}(R)-G(R)\right\} ,  \label{10}
\end{equation}

where $R$, $N$, and $G_{ref}(R)$ stand for the radius of hypersurface $%
\Sigma $, the dimension and an optional non-negative reference function,
respectively. $G_{ref}(R)$ assures zero energy for the spacetime. The
functions of the metric (\ref{9}) read%
\begin{eqnarray}
F(R) &=&\sqrt{f\left( r\right) },  \nonumber \\
G(R) &=&\frac{RF(R)}{2}=\frac{r\sqrt{f\left( r\right) }}{2},  \label{11}
\end{eqnarray}

such that

\begin{equation}
G_{ref}\left( R\right) =G|_{z=0},  \label{12}
\end{equation}

in which $z=r_{h}^{2}/r^{2}.$

After applying these to the solution $f\left( r\right) =1-z$\ and making a
straightforward calculation, the quasilocal mass of the 4-D $Z0$LBH (\ref{7a}%
) can be obtained as%
\begin{equation}
M_{BY}=r_{h}^{2}/4,  \label{13}
\end{equation}

or%
\begin{equation}
r_{h}^{2}=4M_{BY}=4M_{QL.}  \label{14}
\end{equation}

One may check if this result is in consistent with the mass derived from $%
dM=TdS$. One should first calculate the surface gravity \cite{wald0} with
the timelike Killing vector $\xi ^{\mu }=\left( 1,0,0,0\right) $

\begin{equation}
\kappa =\lim_{r\rightarrow r_{h}}\sqrt{\frac{\xi ^{\mu }\nabla _{\mu }\xi
_{\nu }\xi ^{\rho }\nabla _{\rho }\xi ^{\nu }}{-\xi ^{2}}},  \label{15}
\end{equation}

which corresponds to the following expression \cite{huang}

\begin{equation}
\kappa =\frac{r_{h}f^{\prime }(r_{h})}{4}=\frac{1}{2}.  \label{16}
\end{equation}

Therefore, the Hawking temperature \cite{wald0,wald1} reads%
\begin{equation}
T_{H}=\frac{\kappa }{2\pi }=\frac{1}{4\pi }.  \label{17}
\end{equation}

As it can be deduced from that constant Hawking temperature, $Z0$LBH
radiates with isothermal process. After substituting Eq. (17) into $%
dM=T_{H}dS_{BH}$, we derive the entropy as%
\begin{equation}
S_{BH}=\pi r_{h}^{2}=\frac{A_{BH}}{4},  \label{18}
\end{equation}

which agrees with the Bekenstein-Hawking entropy \cite%
{Beks2,Beks3,Beks4,Beks5}.

\section{Dirac Field on $Z0$LBH}

In GR,in order to study the Dirac equation in curved spacetimes, many
formalisms have been developed such as NP and spinor formalisms \cite{newman}%
. In order to solve Dirac equation in $Z0$LBH geometry, we use NP formalism 
\cite{newman,newman2}\ and consider a massive Dirac field \ In order for
solving the Dirac equation, we separate the Dirac equation.and give the
solution of angular equation with the separation constant $\lambda $.
Finally, we show how the radial equation reduces to the ZE \cite{chandra}
with an effective potential in the NH.

The basis vectors of the null tetrad \cite{newman} in the geometry defined
by the line element Eq. (\ref{7a}) are chosen as

\begin{eqnarray}
l_{\mu } &=&\sqrt{\frac{f(r)}{2}}\left[ 1,-\frac{2}{rf\left( r\right) },0,0%
\right] ,  \nonumber \\
n_{\mu } &=&\sqrt{\frac{f(r)}{2}}\left[ 1,\frac{2}{rf\left( r\right) },0,0%
\right] ,  \nonumber \\
m_{\mu } &=&-\frac{r}{\sqrt{2}}\left[ 0,0,1,i\sin \left( \theta \right) %
\right] ,  \nonumber \\
\overline{m}_{\mu } &=&\frac{r}{\sqrt{2}}\left[ 0,0,-1,i\sin \left( \theta
\right) \right] .  \label{19}
\end{eqnarray}

The nonzero NP spin coefficients \cite{newman} regarding the above covariant
null tetrad are found as

\begin{eqnarray}
\alpha &=&-\beta =-\frac{\sqrt{2}}{4}\frac{\cot \left( \theta \right) }{r}, 
\nonumber \\
\varepsilon &=&\gamma =\frac{\sqrt{2}}{16}\frac{r}{\sqrt{f\left( r\right) }}%
\left[ \partial _{r}f\left( r\right) \right] ,  \nonumber \\
\rho &=&\mu =-\frac{\sqrt{2}}{4}\sqrt{f\left( r\right) }.  \label{20}
\end{eqnarray}

In NP formalism, in the case of test ($q=0$)\ Dirac particles, the Dirac
equations can be expressed by the well known Chandrasekhar-Dirac equations
(CDEs) \cite{chandra}\ with the aid of the spin coefficients as follows

\begin{eqnarray}
\left( D+\varepsilon -\rho \right) F_{1}+\left( \overline{\delta }+\pi
-\alpha \right) F_{2} &=&i\mu _{p}G_{1},  \nonumber \\
\left( \Delta +\mu -\gamma \right) F_{2}+\left( \delta +\beta -\tau \right)
F_{1} &=&i\mu _{p}G_{2},  \nonumber \\
\left( D+\overline{\varepsilon }-\overline{\rho }\right) G_{2}-\left( \delta
+\overline{\pi }-\overline{\alpha }\right) G_{1} &=&i\mu _{p}F_{2}, 
\nonumber \\
\left( \Delta +\overline{\mu }-\overline{\gamma }\right) G_{1}-\left( 
\overline{\delta }+\overline{\beta }-\overline{\tau }\right) G_{2} &=&i\mu
_{p}F_{1},  \label{21}
\end{eqnarray}

where $\mu ^{\ast }=\sqrt{2}\mu _{p}$\ is the mass of the uncharged Dirac
particles. The directional derivatives corresponding to the null tetrad are
defined as \cite{chandra}

\begin{eqnarray}
D &=&l^{j}\partial _{j},  \nonumber \\
\Delta &=&n^{j}\partial _{j},  \nonumber \\
\delta &=&m^{j}\partial _{j},  \nonumber \\
\overline{\delta } &=&\overline{m}^{j}\partial _{j},  \label{22}
\end{eqnarray}

where a bar over a quantity denotes complex conjugation.

The wave functions $F_{1}$, $F_{2}$, $G_{1}$, and $G_{2}$ which represent
the Dirac spinors are assumed to be \cite{chandra}

\begin{eqnarray}
F_{1} &=&f_{1}\left( r\right) A_{1}\left( \theta \right) \exp \left[ i\left(
\omega t+m\varphi \right) \right] ,  \nonumber \\
G_{1} &=&g_{1}\left( r\right) A_{2}\left( \theta \right) \exp \left[ i\left(
\omega t+m\varphi \right) \right] ,  \nonumber \\
F_{2} &=&f_{2}\left( r\right) A_{3}\left( \theta \right) \exp \left[ i\left(
\omega t+m\varphi \right) \right] ,  \nonumber \\
G_{2} &=&g_{2}\left( r\right) A_{4}\left( \theta \right) \exp \left[ i\left(
\omega t+m\varphi \right) \right] ,  \label{23}
\end{eqnarray}

where $\omega $\ (positive and real) and\ $m$ are the frequency of the
incoming wave related to the energy of the Dirac particle and the azimuthal
quantum number of the wave, respectively.

Substituting Eq. (\ref{23}) and Eq. (\ref{20}) into Eq. (\ref{21}), one can
simplify the Dirac equations to get

\begin{eqnarray}
\frac{\widetilde{Z}f_{1}\left( r\right) }{f_{2}\left( r\right) }+\frac{%
LA_{3}\left( \theta \right) }{A_{1}\left( \theta \right) }-i\mu ^{\ast }r%
\frac{g_{1}\left( r\right) A_{2}\left( \theta \right) }{f_{2}\left( r\right)
A_{1}\left( \theta \right) } &=&0,  \nonumber \\
\frac{\overline{\widetilde{Z}}f_{2}\left( r\right) }{f_{1}\left( r\right) }-%
\frac{L^{\dagger }A_{1}\left( \theta \right) }{A_{3}\left( \theta \right) }%
+i\mu ^{\ast }r\frac{g_{2}\left( r\right) A_{4}\left( \theta \right) }{%
f_{1}\left( r\right) A_{3}\left( \theta \right) } &=&0,  \nonumber \\
\frac{\widetilde{Z}g_{2}\left( r\right) }{g_{1}\left( r\right) }-\frac{%
L^{\dagger }A_{2}\left( \theta \right) }{A_{4}\left( \theta \right) }-i\mu
^{\ast }r\frac{f_{2}\left( r\right) A_{3}\left( \theta \right) }{g_{1}\left(
r\right) A_{4}\left( \theta \right) } &=&0,  \nonumber \\
\frac{\overline{\widetilde{Z}}g_{1}\left( r\right) }{g_{2}\left( r\right) }+%
\frac{LA_{4}\left( \theta \right) }{A_{2}\left( \theta \right) }+i\mu ^{\ast
}r\frac{f_{1}\left( r\right) A_{1}\left( \theta \right) }{g_{2}\left(
r\right) A_{2}\left( \theta \right) } &=&0,  \label{24}
\end{eqnarray}

with the radial and the angular operators, respectively

\begin{eqnarray}
\widetilde{Z} &=&\frac{i\omega r}{\sqrt{f\left( r\right) }}+\frac{1}{2}r^{2}%
\sqrt{f\left( r\right) }\partial _{r}+\frac{1}{8}\frac{r^{2}}{\sqrt{f\left(
r\right) }}\left[ \partial _{r}f\left( r\right) \right] +\frac{1}{2}r\sqrt{%
f\left( r\right) },  \nonumber \\
\overline{\widetilde{Z}} &=&-\frac{i\omega r}{\sqrt{f\left( r\right) }}+%
\frac{1}{2}r^{2}\sqrt{f\left( r\right) }\partial _{r}+\frac{1}{8}\frac{r^{2}%
}{\sqrt{f\left( r\right) }}\left[ \partial _{r}f\left( r\right) \right] +%
\frac{1}{2}r\sqrt{f\left( r\right) },  \label{25}
\end{eqnarray}

and

\begin{eqnarray}
L &=&\partial _{\theta }+\frac{m}{\sin \left( \theta \right) }+\frac{\cot
\left( \theta \right) }{2},  \nonumber \\
L^{\dagger } &=&\partial _{\theta }-\frac{m}{\sin \left( \theta \right) }+%
\frac{\cot \left( \theta \right) }{2}.  \label{26}
\end{eqnarray}

As it is obvious from Eq. (\ref{24}) that $\left\{ f_{1}\text{, }f_{2}\text{%
, }g_{1}\text{, }g_{2}\right\} $ and $\left\{ A_{1}\text{, }A_{2}\text{, }%
A_{3}\text{, }A_{4}\right\} $ are the functions of two distinct variables $r$%
\ and $\theta $,respectively, one can introduce a separation constant $%
\lambda $ and assume that

\begin{eqnarray}
f_{1}\left( r\right) &=&g_{2}\left( r\right) ,  \nonumber \\
f_{2}\left( r\right) &=&g_{1}\left( r\right) ,  \nonumber \\
A_{1}\left( \theta \right) &=&A_{2}\left( \theta \right) ,  \nonumber \\
A_{3}\left( \theta \right) &=&A_{4}\left( \theta \right) .  \label{27}
\end{eqnarray}

to split Eq. (\ref{24}) into two sets of radial and angular equations

\begin{eqnarray}
\overline{\widetilde{Z}}g_{1}\left( z\right) &=&\left( \lambda -i\mu ^{\ast
}r\right) g_{2}\left( z\right) ,  \nonumber \\
\widetilde{Z}g_{2}\left( z\right) &=&\left( \lambda +i\mu ^{\ast }r\right)
g_{1}\left( z\right) ,  \label{28}
\end{eqnarray}

and

\begin{eqnarray}
L^{\dagger }A_{1}\left( \theta \right) &=&\lambda A_{3}\left( \theta \right)
,  \nonumber \\
LA_{3}\left( \theta \right) &=&-\lambda A_{1}\left( \theta \right) .
\label{29}
\end{eqnarray}

At this point it would me more convenient to make the choice of massless
Dirac particles ($\mu ^{\ast }=0$) to deal with above sets of equations.

In the spherical case, the angular operators (or the so-called laddering
operators) $L^{\dagger }$ and $L$\ lead the spin weighted spheroidal
harmonics $_{s}Y_{l^{\shortmid }}^{m}\left( \theta \right) $ \cite%
{castillo,newman2,newman,berti,Teukolsky} and they are governed by

\begin{eqnarray}
\left( \partial _{\theta }-\frac{m}{\sin \theta }-s\cot \theta \right) \left[
_{s}Y_{l^{\shortmid }}^{m}\left( \theta \right) \right] &=&-\sqrt{\left(
l^{\shortmid }-s\right) \left( l^{\shortmid }+s+1\right) }\left[
_{s+1}Y_{l^{\shortmid }}^{m}\left( \theta \right) \right] ,  \nonumber \\
\left( \partial _{\theta }+\frac{m}{\sin \theta }+s\cot \theta \right) \left[
_{s}Y_{l^{\shortmid }}^{m}\left( \theta \right) \right] &=&\sqrt{\left(
l^{\shortmid }+s\right) \left( l^{\shortmid }-s+1\right) }\left[
_{s-1}Y_{l^{\shortmid }}^{m}\left( \theta \right) \right] .  \label{30}
\end{eqnarray}

The spin weighted spheroidal harmonics $_{s}Y_{l^{\shortmid }}^{m}\left(
\theta \right) $ has the following generic form \cite{Teukolsky}

\begin{eqnarray}
_{s}Y_{l^{\shortmid }}^{m}\left( \theta ,\varphi \right) &=&\exp \left(
im\varphi \right) \sqrt{\frac{2l^{\shortmid }+1}{4\pi }\frac{\left(
l^{\shortmid }+m\right) !\left( l^{\shortmid }-m\right) !}{\left(
l^{\shortmid }+s\right) !\left( l^{\shortmid }-s\right) !}}\left[ \sin
\left( \frac{\theta }{2}\right) \right] ^{2l^{\shortmid }}  \nonumber \\
&&\times \sum\limits_{r=-l^{\shortmid }}^{l^{\shortmid }}\left( -1\right)
^{l^{\shortmid }+m-r}\left( 
\begin{array}{c}
l^{\shortmid }-s \\ 
r-s%
\end{array}%
\right) \left( 
\begin{array}{c}
l^{\shortmid }+s \\ 
r-m%
\end{array}%
\right) \left[ \cot \left( \frac{\theta }{2}\right) \right] ^{2r-m-s}
\label{31}
\end{eqnarray}

where $l^{\shortmid }$ is the angular quantum number and $s$\ is the spin
weight with $l^{\shortmid }=\left\vert s\right\vert ,$ $\left\vert
s\right\vert +1,$ $\left\vert s\right\vert +2...$ and $-l^{\shortmid
}<m<+l^{\shortmid }$. Thus, having $s=\pm \frac{1}{2}$, and comparing Eq. (%
\ref{29}) with Eq. (\ref{30}) one can identify

\begin{eqnarray}
A_{1}\left( \theta \right) &=&_{-\frac{1}{2}}Y_{l^{\shortmid }}^{m}\left(
\theta \right) ,  \nonumber \\
A_{3}\left( \theta \right) &=&_{\frac{1}{2}}Y_{l^{\shortmid }}^{m}\left(
\theta \right) ,  \label{32}
\end{eqnarray}

and obtain the separation constant $\lambda $ which is the eigenvalue of the
spin weighted spheroidal harmonic \cite{Teukolsky} equation as

\begin{equation}
\lambda =-\left( l^{\shortmid }+\frac{1}{2}\right) .  \label{33}
\end{equation}

\subsection{Effective Potential}

Now, we give the effective potential and obtain the solution to the radial
part of the Dirac equation (\ref{28}). In fact, it is appropriate to alter
the radial equations in a Schr\"{o}dinger-type with an effective potential,
in order to investigate the BQNMs \cite{BQNM1,BQNM2,BQNM3,Hod}.

Defining a new function in terms of the metric function

\begin{equation}
Y\left( r\right) =\frac{1}{8}\frac{r^{2}}{\sqrt{f\left( r\right) }}\partial
_{r}f\left( r\right) +\frac{1}{2}r\sqrt{f\left( r\right) }  \label{34}
\end{equation}

and substituting the scalings

\begin{eqnarray}
g_{1}\left( r\right) &=&G_{1}\left( r\right) \exp \left[ -2\int \frac{%
Y\left( r\right) }{r^{2}\sqrt{f\left( r\right) }}dr\right] ,  \nonumber \\
g_{2}\left( r\right) &=&G_{2}\left( r\right) \exp \left[ -2\int \frac{%
Y\left( r\right) }{r^{2}\sqrt{f\left( r\right) }}dr\right] ,  \label{35}
\end{eqnarray}

into the radial equations Eq. (\ref{28})\ and setting $\mu ^{\ast }=0$, one
gets

\begin{eqnarray}
\left[ A\left( r\right) \partial _{r}-i\omega \right] G_{1}\left( r\right)
&=&B\left( r\right) G_{2}\left( r\right) ,  \nonumber \\
\left[ A\left( r\right) \partial _{r}+i\omega \right] G_{2}\left( r\right)
&=&B\left( r\right) G_{1}\left( r\right) ,  \label{36}
\end{eqnarray}

where the functions $A(r)$ and $B(r)$ read

\begin{equation}
A\left( r\right) =\frac{rf\left( r\right) }{2},  \label{37}
\end{equation}

and

\begin{equation}
B\left( r\right) =\frac{\lambda \sqrt{f\left( r\right) }}{r},  \label{37b}
\end{equation}

respectively. Introducing the tortoise coordinate as

\begin{equation}
dr_{\ast }=\frac{1}{A\left( r\right) }dr,  \label{38}
\end{equation}

leads us to express Eq. (\ref{36}) as follows

\begin{eqnarray}
\left[ \partial _{r_{\ast }}-i\omega \right] G_{1}\left( r_{\ast }\right)
&=&B\left( r\right) G_{2}\left( r_{\ast }\right) ,  \nonumber \\
\left[ \partial _{r_{\ast }}+i\omega \right] G_{2}\left( r_{\ast }\right)
&=&B\left( r\right) G_{1}\left( r_{\ast }\right) .  \label{39}
\end{eqnarray}

In order to decouple the above equations, we consider the solutions of the
form

\begin{eqnarray}
G_{1}\left( r_{\ast }\right) &=&P_{1}\left( r_{\ast }\right) +P_{2}\left(
r_{\ast }\right) ,  \nonumber \\
G_{2}\left( r_{\ast }\right) &=&P_{1}\left( r_{\ast }\right) -P_{2}\left(
r_{\ast }\right) ,  \label{40}
\end{eqnarray}

which yields the two radial equations of Schr\"{o}dinger-type (or the
so-called ZE \cite{chandra})

\begin{eqnarray}
\partial _{r_{\ast }}^{2}P_{1}\left( r_{\ast }\right) +\left[ \omega
^{2}-V_{1}\right] P_{1}\left( r_{\ast }\right) &=&0,  \nonumber \\
\partial _{r_{\ast }}^{2}P_{2}\left( r_{\ast }\right) +\left[ \omega
^{2}-V_{2}\right] P_{2}\left( r_{\ast }\right) &=&0,  \label{41}
\end{eqnarray}

whose associated Zerilli potentials are given by

\begin{eqnarray}
V_{1} &=&B^{2}\left( r\right) +\partial _{r_{\ast }}B\left( r\right) =\frac{%
\lambda \sqrt{r^{2}-r_{h}^{2}}}{2r^{4}}\left( 2\lambda \sqrt{r^{2}-r_{h}^{2}}%
+2r_{h}^{2}-r^{2}\right) ,  \nonumber \\
V_{2} &=&B^{2}\left( r\right) -\partial _{r_{\ast }}B\left( r\right) =\frac{%
\lambda \sqrt{r^{2}-r_{h}^{2}}}{2r^{4}}\left( 2\lambda \sqrt{r^{2}-r_{h}^{2}}%
-2r_{h}^{2}+r^{2}\right) .  \label{42}
\end{eqnarray}

For studying the fermion BQNMs analytically, we follow a recent study of 
\cite{Hod}. Taking

\begin{equation}
r=xr_{h}+r_{h},  \label{43}
\end{equation}

and substituting into Eq. (\ref{37}) gives

\begin{equation}
A\left( x\right) \mathbf{=}r_{h}x\frac{\left( x+2\right) }{2\left(
x+1\right) },  \label{44}
\end{equation}

with its NH form

\begin{equation}
A_{NH}\left( x\right) \approx r_{h}x+O\left( x^{2}\right) =2\kappa r_{h}x.
\label{45}
\end{equation}

This allows us to write the NH forms of the Zerilli potentials (\ref{42})\ as

\begin{eqnarray}
V_{1}^{NH}\left( x\right) &=&\frac{\lambda }{\sqrt{2}r_{h}}\sqrt{x}+\frac{%
2\lambda ^{2}}{r_{h}^{2}}x+O\left( x^{3/2}\right) ,  \nonumber \\
V_{2}^{NH}\left( x\right) &=&-\frac{\lambda }{\sqrt{2}r_{h}}\sqrt{x}+\frac{%
2\lambda ^{2}}{r_{h}^{2}}x+O\left( x^{3/2}\right) .  \label{46}
\end{eqnarray}

Setting a new coordinate

\begin{equation}
y\mathbf{=}\int \frac{\kappa r_{h}}{A_{NH}\left( x\right) }dx\mathbf{=}%
\frac{1}{2}\ln \left( x\right) =\kappa r_{\ast },  \label{47}
\end{equation}

with its limits

\begin{eqnarray}
\lim\limits_{r\rightarrow r_{h}}y &=&-\infty ,  \nonumber \\
\lim\limits_{r\rightarrow \infty }y &=&\infty ,  \label{48}
\end{eqnarray}

one can express the radial coordinate $x$ in terms of the surface gravity as

\begin{equation}
x=\exp \left( 2y\right) =\exp \left( 2\kappa r_{\ast }\right) .  \label{49}
\end{equation}

Therefore, we can recast the NH Zerilli potentials (or the so-called
effective potentials) (\ref{46}), in the leading order terms as follows

\begin{eqnarray}
V_{1}^{NH}\left( y\right) &=&\frac{\lambda }{\sqrt{2}r_{h}}\exp \left(
y\right) ,  \nonumber \\
V_{2}^{NH}\left( y\right) &=&-\frac{\lambda }{\sqrt{2}r_{h}}\exp \left(
y\right) .  \label{50}
\end{eqnarray}

The NH ZEs in which $\widetilde{\omega }=\omega /\kappa $

\begin{eqnarray}
\frac{d^{2}}{dy^{2}}P_{1}(y)+\left( \widetilde{\omega }^{2}-\frac{%
V_{1}^{NH}\left( y\right) }{\kappa ^{2}}\right) P_{1}(y) &=&0,  \nonumber \\
\frac{d^{2}}{dy^{2}}P_{2}(y)+\left( \widetilde{\omega }^{2}-\frac{%
V_{2}^{NH}\left( y\right) }{\kappa ^{2}}\right) P_{2}(y) &=&0,  \label{51}
\end{eqnarray}

can be solved after inserting the value of the separation constant. The
solutions of these equations are obtained in terms of the Bessel functions
of the first and the second kind \cite{Abramowitz} as follows

\begin{eqnarray}
P_{j}\left( y\right) &=&C_{j1}J_{-2i\widetilde{\omega }}\left[ \aleph 2^{5/4}%
\sqrt{\frac{2l^{\shortmid }+1}{r_{h}}}\exp \left( y/2\right) \right]+\nonumber \\
&&C_{j2}Y_{-2i\widetilde{\omega }}\left[ \aleph 2^{5/4}\sqrt{\frac{%
2l^{\shortmid }+1}{r_{h}}}\exp \left( y/2\right) \right] ,  \label{52}
\end{eqnarray}

where $C_{j1}$ and $C_{j2}$ are constants and

\begin{equation}
\aleph =\left( i\right) ^{j-1},\text{ \ \ \ \ }j=1,2.  \label{53}
\end{equation}

\section{BQNMs and Quantum Spectra of $Z0$LBH}

In this section, our interest is to compute the Dirac BQNMs. First, we
impose the purely ingoing wave condition at the event horizon for QNMs to
appear \cite{chandra}. Then, we impose the DBC and the NBC in order for
getting the resonance conditions. At this point, we use an iteration in
order to solve the resonance conditions \cite{Hod,Sakalli}.

The solutions of the NH ZEs can be rewritten as

\begin{equation}
P_{j}\left( x\right) =C_{j1}J_{-2i\widetilde{\omega }}\left( 4\aleph \sqrt{%
\Omega \sqrt{x}}\right) +C_{j2}Y_{-2i\widetilde{\omega }}\left( 4\aleph 
\sqrt{\Omega \sqrt{x}}\right)  \label{54}
\end{equation}

with the parameter

\begin{equation}
\Omega =\frac{2l^{\shortmid }+1}{2\sqrt{2}r_{h}}.  \label{55}
\end{equation}

Using the limiting forms of the Bessel functions \cite{Abramowitz,Olver}
given by

\begin{eqnarray}
J_{\nu }\left( z\right) &\sim &\frac{\left[ \left( 1/2\right) z\right] ^{\nu
}}{\Gamma \left( 1+\nu \right) },\text{ \ \ \ \ }\nu \neq -1,-2,-3,... 
\nonumber \\
Y_{\nu }\left( z\right) &\sim &-\frac{1}{\pi }\Gamma \left( \nu \right)
\left( \frac{1}{2}z\right) ^{-\nu },\text{ \ \ \ \ }\Re \nu >0  \label{56}
\end{eqnarray}

the NH ($\exp (y/2)\ll 1$) behavior of the solution can be obtained as

\begin{eqnarray}
P_{j} &\sim &C_{j1}\frac{\left( 2\aleph \sqrt{\Omega }\right) ^{-2i%
\widetilde{\omega }}}{\Gamma \left( 1-2i\widetilde{\omega }\right) }\exp (-i%
\widetilde{\omega }y)-C_{j2}\frac{1}{\pi }\Gamma \left( -2i\widetilde{\omega 
}\right) \left( 2\aleph \sqrt{\Omega }\right) ^{2i\widetilde{\omega }}\exp (i%
\widetilde{\omega }y)  \nonumber \\
&=&C_{j1}\frac{\left( 2\aleph \sqrt{\Omega }\right) ^{-2i\widetilde{\omega }}%
}{\Gamma \left( 1-2i\widetilde{\omega }\right) }\exp (-i\omega r_{\ast
})-C_{j2}\frac{1}{\pi }\Gamma \left( -2i\widetilde{\omega }\right) \left(
2\aleph \sqrt{\Omega }\right) ^{2i\widetilde{\omega }}\exp (i\omega r_{\ast
}).  \label{57}
\end{eqnarray}

Imposing the boundary condition at the event horizon for BQNMs requires us
to vanish the outgoing waves by choosing $C_{j2}=0$. Therefore the proper
solution of Eq. (\ref{51}) becomes

\begin{equation}
P_{j}\left( x\right) =C_{j1}J_{-2i\widetilde{\omega }}\left( 4\aleph \sqrt{%
\Omega \sqrt{x}}\right) .  \label{58}
\end{equation}

Taking into account the DBC at the horizon (confining cage) \cite%
{Hod,Sakalli,okawa,gul}

\begin{equation}
P_{j}\left( x\right) |_{x=x_{m}}=0,  \label{59}
\end{equation}

one gets the condition

\begin{equation}
J_{-2i\widetilde{\omega }}\left( 4\aleph \sqrt{\Omega \sqrt{x_{m}}}\right)
=0.  \label{60}
\end{equation}

With the aid of the relation \cite{Abramowitz},

\begin{equation}
Y_{\nu }\left( z\right) =J_{\nu }\left( z\right) \cot \left( \nu \pi \right)
-J_{-\nu }\left( z\right) \csc \left( \nu \pi \right)  \label{61}
\end{equation}

the boundary condition Eq. (\ref{60}) can be stated as

\begin{equation}
\tan \left( 2i\widetilde{\omega }\pi \right) =\frac{J_{2i\widetilde{\omega }%
}\left( 4\aleph \sqrt{\Omega \sqrt{x_{m}}}\right) }{Y_{2i\widetilde{\omega }%
}\left( 4\aleph \sqrt{\Omega \sqrt{x_{m}}}\right) }  \label{62}
\end{equation}

The boundary of the cage is at the event horizon \ Thus, one can use the
limiting forms of the Bessel functions \cite{Abramowitz} in the above
condition and obtain the resonance condition as follows

\begin{eqnarray}
\tan \left( 2i\widetilde{\omega }\pi \right) &\sim &-\frac{\pi \left(
2\aleph \sqrt{\sqrt{z_{m}}}\right) ^{4i\widetilde{\omega }}}{\Gamma \left( 2i%
\widetilde{\omega }\right) \Gamma \left( 2i\widetilde{\omega }+1\right) }=i%
\frac{\pi (\aleph ^{2})^{2i\widetilde{\omega }}}{2\widetilde{\omega }\Gamma
^{2}\left( 2i\widetilde{\omega }\right) }\left( 4\sqrt{z_{m}}\right) ^{2i%
\widetilde{\omega }}  \nonumber \\
&=&i\frac{\pi \exp \left( -4\pi \widetilde{\omega }/j\right) }{2\widetilde{%
\omega }\Gamma ^{2}\left( 2i\widetilde{\omega }\right) }\left( 4\sqrt{z_{m}}%
\right) ^{2i\widetilde{\omega }},  \label{63}
\end{eqnarray}

where $z_{m}=\Omega ^{2}x_{m}$.

Imposing the NBC \cite{Hod,Sakalli,okawa,gul}\ given by

\begin{equation}
\frac{dR(x)}{dx}|_{x=x_{m}}=0,  \label{64}
\end{equation}

we find

\begin{equation}
J_{-2i\widetilde{\omega }-1}\left( 4\aleph \sqrt{\sqrt{z_{m}}}\right) -J_{-2i%
\widetilde{\omega }+1}\left( 4\aleph \sqrt{\sqrt{z_{m}}}\right) =0
\label{65}
\end{equation}

from Eq. (\ref{61}) we obtain the following relation

\begin{equation}
Y_{\nu +1}\left( z\right) -Y_{\nu -1}\left( z\right) =\cot (\nu \pi )\left[
J_{\nu +1}\left( z\right) -J_{\nu -1}\left( z\right) \right] -\csc (\nu \pi )%
\left[ J_{-\nu -1}\left( z\right) -J_{-\nu +1}\left( z\right) \right]
\label{66}
\end{equation}

We combine the equations (\ref{65}) and (\ref{66}) and express the NBC's
resonance condition as

\begin{equation}
\tan (2i\widetilde{\omega }\pi )=\frac{J_{2i\widetilde{\omega }-1}\left(
4\aleph \sqrt{\sqrt{z_{m}}}\right) }{Y_{2i\widetilde{\omega }+1}\left(
4\aleph \sqrt{\sqrt{z_{m}}}\right) }\left[ \frac{-1+J_{2i\widetilde{\omega }%
+1}\left( 4\aleph \sqrt{\sqrt{z_{m}}}\right) /J_{2i\widetilde{\omega }%
-1}\left( 4\aleph \sqrt{\sqrt{z_{m}}}\right) }{1-Y_{2i\widetilde{\omega }%
-1}\left( 4\aleph \sqrt{\sqrt{z_{m}}}\right) /Y_{2i\widetilde{\omega }%
+1}\left( 4\aleph \sqrt{\sqrt{z_{m}}}\right) }\right] .  \label{67}
\end{equation}

Using the limiting forms (\ref{56}), one finds

\begin{equation}
\frac{J_{2i\widetilde{\omega }+1}\left( 4\aleph \sqrt{\sqrt{z_{m}}}\right) }{%
J_{2i\widetilde{\omega }-1}\left( 4\aleph \sqrt{\sqrt{z_{m}}}\right) }\equiv 
\frac{Y_{2i\widetilde{\omega }-1}\left( 4\aleph \sqrt{\sqrt{z_{m}}}\right) }{%
Y_{2i\widetilde{\omega }+1}\left( 4\aleph \sqrt{\sqrt{z_{m}}}\right) }\sim
O(z_{m}),  \label{68}
\end{equation}

and reads the resonance condition (\ref{67}) in the NH as

\begin{eqnarray}
\tan (2i\widetilde{\omega }\pi ) &\sim &-\frac{J_{2i\widetilde{\omega }%
-1}\left( 4\aleph \sqrt{\sqrt{z_{m}}}\right) }{Y_{2i\widetilde{\omega }%
+1}\left( 4\aleph \sqrt{\sqrt{z_{m}}}\right) }=-i\frac{\pi (\aleph ^{2})^{2i%
\widetilde{\omega }}}{2\widetilde{\omega }\Gamma ^{2}\left( 2i\widetilde{%
\omega }\right) }\left( 4\sqrt{z_{m}}\right) ^{2i\widetilde{\omega }} 
\nonumber \\
&=&-i\frac{\pi \exp \left( -4\pi \widetilde{\omega }/j\right) }{2\widetilde{%
\omega }\Gamma ^{2}\left( 2i\widetilde{\omega }\right) }\left( 4\sqrt{z_{m}}%
\right) ^{2i\widetilde{\omega }}.  \label{69}
\end{eqnarray}

To solve the resonance conditions, we use an iteration method \cite%
{Hod,Sakalli,gul}, since the obtained resonance conditions are small
quantities. The zeroth order resonance condition \cite{Hod,Sakalli,gul} has
the form

\begin{equation}
\tan (2i\widetilde{\omega }_{n}^{(0)}\pi )=0,  \label{70}
\end{equation}

which means that

\begin{equation}
\widetilde{\omega }_{n}^{(0)}=-i\frac{n}{2},\text{ \ \ \ \ }(n=0,1,2,...).
\label{71}
\end{equation}

The first order resonance condition can be obtained ,by substituting Eq. (%
\ref{71}) into the r.h.s. of (\ref{63}) and (\ref{69}), as follows

\begin{equation}
\tan (2i\widetilde{\omega }_{n}^{(1)}\pi )=\pm i\frac{\pi \exp \left( 2i\pi
n/j\right) }{\left( -in\right) \Gamma ^{2}\left( n\right) }\left( 4\sqrt{%
z_{m}}\right) ^{n},  \label{72}
\end{equation}

which reduces to

\begin{equation}
\tan (2i\widetilde{\omega }_{n}^{(1)}\pi )=\mp n\frac{\pi }{\left( n!\right)
^{2}}\left[ \left( -1\right) ^{2/j}4\sqrt{z_{m}}\right] ^{n}.  \label{73}
\end{equation}

Using $\tan \left( x+n\pi \right) =\tan \left( x\right) \approx x$ for $x\ll 
$ yields

\begin{equation}
2i\widetilde{\omega }_{n}\pi =n\pi \left\{ 1\mp \frac{1}{\left( n!\right)
^{2}}\left[ \left( -1\right) ^{2/j}4\sqrt{z_{m}}\right] ^{n}\right\}
\label{74}
\end{equation}

Hence we find

\begin{equation}
\widetilde{\omega }_{n}=-i\frac{n}{2}\left\{ 1\mp \frac{1}{\left( n!\right)
^{2}}\left[ \left( -1\right) ^{2/j}4\sqrt{z_{m}}\right] ^{n}\right\} ,
\label{75}
\end{equation}

and read the Dirac BQNMs as follows

\begin{equation}
\omega _{n}=-i\kappa \frac{n}{2}\left\{ 1\mp \frac{1}{\left( n!\right) ^{2}}%
\left[ \left( -1\right) ^{2/j}4\sqrt{z_{m}}\right] ^{n}\right\} ,\text{ \ \
\ \ }(n=0,1,2,...).  \label{76}
\end{equation}

where $n$\ stands for the overtone quantum number (resonance parameter) \cite%
{Hod2}.

For the highly excited states ($n\longrightarrow \infty $), BQNM frequencies
read

\begin{equation}
\omega _{n}\approx -i\kappa \frac{n}{2},\text{ \ \ \ \ }(n\longrightarrow
\infty ).  \label{77}
\end{equation}

The transition frequency from MM \cite{maggiore} can be obtained as

\begin{equation}
\Delta \omega _{I}=\frac{\kappa }{2}=\frac{\pi T_{H}}{\hbar }.  \label{78}
\end{equation}

Therefore, the adiabatic invariant quantity \cite{medved,vagenas,Sakalli2}\
becomes

\begin{equation}
I_{adb}=\frac{\hbar }{\pi }S^{BH}.  \label{79}
\end{equation}

With Bohr-Sommerfeld quantization rule ($I_{adb}=n\hbar $) \cite{BS}, the
entropy spectrum of $Z0$LBHs is determined as

\begin{equation}
S_{n}^{BH}=\pi n,  \label{80}
\end{equation}

and using $S_{n}^{BH}=A_{n}^{BH}/4\hbar $ one may obtain the area spectrum

\begin{equation}
A_{n}^{BH}=4\pi \hbar n,  \label{81}
\end{equation}

with the minimum spacing given by

\begin{equation}
\Delta A_{\min }=4\pi \hbar .
\end{equation}

As such, one concludes that the entropy/area spectra of the $Z0$LBHs are
evenly spaced and are independent from the BH parameters whereas the spacing
coefficient reads $\epsilon =4\pi $, which is half the Bekenstein's
result.for the Schwarzschild BH \cite{vagenas,maggiore,medved}.

\section{Discussion}

In the present study, based on the adiabatic invariant formulation in MM
(3),the quantum entropy/area spectra of the $Z0$LBH are investigated. The
massless Dirac equation for the test particles on a $Z0$LBH is solved by the
separation of variables and it is decoupled with an eigenvalue $\lambda $\
into the radial and the angular equations. Particularly, we have obtained
the ZEs with their associated potentials and limited them to the NH region
in order to show analytically the existence of the BQNMs in the presence of
a Dirac field on $Z0$LBH by using an iteration method. For this purpose, we
have considered a confining mirror that is placed in the NH region of the $%
Z0 $LBH. We have therefore showed that the Dirac BQNMs are purely imaginary
and negative which guarantees the stability of these BHs under massless
fermionic field perturbations. After imposing the appropriate boundary
conditions, we have computed the resonant frequencies of the $Z0$LBH which
is confined in a finite-volume cavity (mirror). Later on, MM is applied to
the highly damped Dirac BQNMs to derive the entropy/area spectra of the $Z0$%
LBH which is found to be equally spaced and independent of the BH parameters
as stated in \cite{kothawala,wei,Beks1,Beks2,Beks3,Beks4,Beks5}, although
the spacing coefficient is half of the Bekenstein's original result \cite%
{vagenas,maggiore,medved}.

The QNMs of z=0 and z=2 LBHs were previously studied by \cite{catalan} and 
\cite{gulzzlbh}, respectively.Our solution enables us to investigate a
similar results with the Dirac fields.

Extremal BHs are believed to have connection with the ground states of
quantum gravity which indicates the significance of the spin- 1/2 particles
on such backgrounds. Therefore, the analysis of a Dirac field interacting
with a rotating or a charged rotating BH would be our next interest to
obtain the possible Dirac BQNMs.

\end{document}